\documentclass{article}

\usepackage[english]{babel}

\usepackage[letterpaper,top=2cm,bottom=2cm,left=3cm,right=3cm,marginparwidth=1.75cm]{geometry}

\usepackage{amsmath}
\usepackage{amssymb}
\usepackage{graphicx}
\usepackage[colorlinks=true, allcolors=blue]{hyperref}

\title{Optimal Small-Bitwidth Moduli Set for \\ Residue Number Systems}
\date{}
\author{Danila Gorodecky (Gorodetsky)\\ INESC-ID, University of Lisbon, Portugal \\ \textit{danila.gorodecky@gmail.com}}

\begin{document}
\maketitle

\begin{abstract}
This technical note presents a algorithmic approach for generating optimal sets of co-prime moduli within specified integer ranges. The proposed method addresses the challenge of balancing moduli bit-lengths while maximizing the dynamic range in Residue Number System (RNS) implementations. Experimental results demonstrate that the generated moduli sets achieve optimal dynamic range coverage while maintaining balanced bit-length distribution, making them particularly suitable for parallel hardware implementations based on RNS.
\end{abstract}

\section{Introduction}
The Chinese Reminder Theorem (CRT)~\cite{di_pe_sa} states that there is a one-to-one correspondence between a set of residues $S_1, S_2, ..., S_n$ and an integer number in the range from $0$ to $p_1 \cdot p_2 \cdot ... \cdot p_n - 1 = P - 1$, where $\{p_i|_{1 \leq i \leq n}\}$ is the moduli set of co-prime numbers for which the $\{S_i|_{1 \leq i \leq n}\}$ have been calculated. Since the value of the represented number is invariant under any permutation of the residues, RNS is a non-positional number system.

In comparison to positional number systems, RNS alternative number system has pros and cons. If from one side RNS allows parallelizing computations and hence speeds them up, on the other side it does not allow for directly comparing the magnitude of two numbers represented by their residues.
RNS exhibits data parallelism, computer arithmetic is performed independently on each residue, constituting what is called a channel. 
Since the residues have a lower number of digits than the original binary representation, and arithmetic operations over the residues are performed separately for each modulus of the set, it results in faster computation. RNS is suitable for arithmetic operations such as addition, subtraction, and multiplication, with an arbitrary number of operands.

\begin{figure*}
\centering
    \includegraphics[width=310pt]{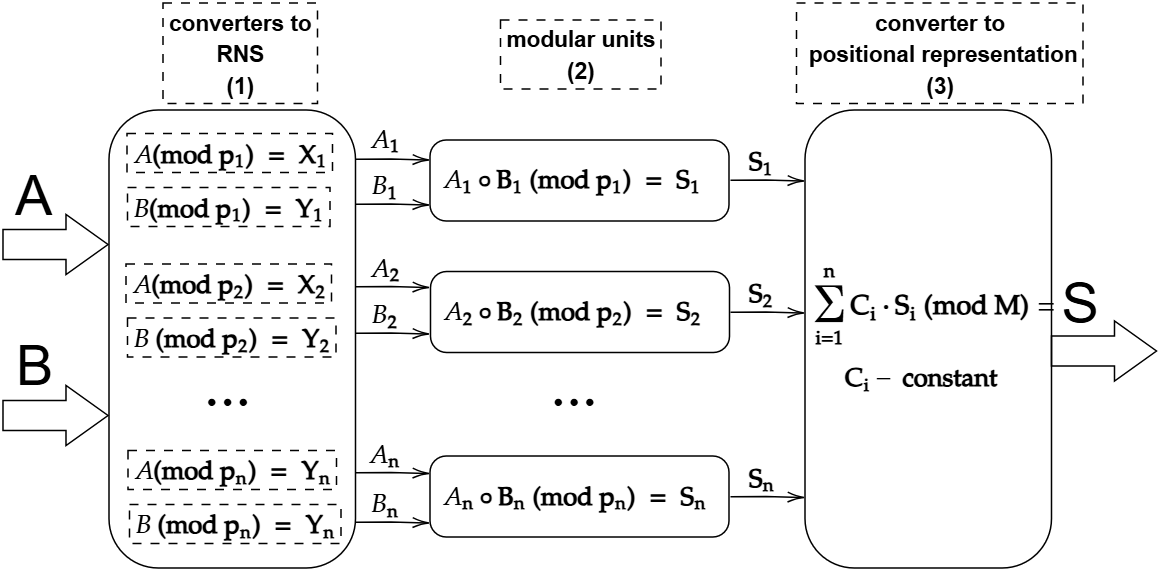}
    \caption{Schematic representation of RNS, where $\circ \in \{+, -, \times\}$.}\label{fig1}
\end{figure*}

Data processing in RNS typically includes the following steps. If the input operands $A_1, A_2, \dots, A_n$ are represented in binary, they have to be converted to modular representations RNS by computing the remainders (or residues) with respect to the moduli $\{p_1, p_2, \dots, p_n\}$; then arithmetic operations are applied to the residues for each modulo $p_i$; finally, the results $S_1,S_2,...,S_n$ might have to be converted back from residues to the binary representation $S$, as depicted in Fig.~\ref{fig1}.

In Fig.~\ref{fig1} block (1) operands are converted into $X (mod\ P)$; in block (2) modular summation, multiplication, and combinations of these operators, such as $A \cdot B + C$ are performed; in block (3) the polynomial form $S_1 \cdot C_1 + S_2 \cdot C_2 + \cdots + S_n \cdot C_n - P \cdot k$ is computed, where $S_1, \cdots S_n$ are outputs of the previous block, $C_1,\cdots, C_n$ are pre-calculated constants, $k$ is a constant which is obtained during the computation of the polynomial and $P=p_1\cdot p_2\cdot ... \cdot p_m$. 
Therefore, the main arithmetic operations needed for RNS computations are the modulo function $X (mod\ P)$, modular summation, and modular multiplication.


For  converting the representation of $S$ from RNS to binary, the CRT and the Mixed Radix Conversion (MRC)~\cite{om_pr, sza_tan} can be applied:
\begin{equation}\label{eq_1}
\begin{split}
S = & (S_1 \cdot C_1 + S_2 \cdot C_2 + ... + S_n \cdot C_n)\:  mod \: P \\
= & S_1 \cdot C_1 + S_2 \cdot C_2 + ... + S_n \cdot C_n - k\cdot P,
\end{split}
\end{equation}
where $k$ is a natural number and $C_i = \frac{P} {p_i} \cdot q_i$, with $q_i \cdot \frac{P} {p_i} (mod \: p_i) = 1$, for $i=1,2,...,n$.


Some drawbacks prevent using RNS for general-purpose computing. For example, it is hard to compute division and comparison in the RNS domain, and binary-RNS forward and backward conversions are overheads in cost and performance. The backward conversion consists of multipliers and adders, according to~\eqref{eq_1}, and either the reduction modulo $P$ is computed or a comparison with the value $k\cdot P$ is performed several times in order to find $k$.
In practice, these problems associated with RNS have been mitigated by adopting small sets of special moduli, which limits the applicability of RNS. Special moduli define specific forms, such as Mersenne primes ($2^k - 1$), Fermat numbers ($2^{2^{k}} + 1$), Cullen numbers ($n\cdot 2^k +1$), Sophie Germain primes $k$, for $2\cdot k + 1$ prime, and other variations of "power of two" numbers~\cite{om_pr,moh,sou_ant,ban_gbo,nan_re}. Special features of these types of moduli simplify the reduction  to calculate the residues. But it is not known any efficient common approach to calculating the residue for an arbitrary divider.

Flexibility is improved by allowing arbitrary modulo sets, which may reduce significantly the bit-width of the residues in comparison to special moduli approaches. For instance, considering  multiplication for 50-bit range in RNS, for a special moduli set may require 17-bit ($2^{17}-1,2^{17},2^{17}+1$) or 6-bit for an arbitrary moduli set ($\{37,39,41,43,47,49,55,59,61\}$). A slower multiplication of 17-bit for the first case is replaced by a faster 6-bit for the second. On the other hand, the difficulty  of calculating the residues for arbitrary moduli sets may overshadow the advantages of multiplication. RNS deals with special values of moduli due also to the complexity of reverse conversion and reduction for arbitrary modulo values.

Despite extensive research in RNS architectures, as comprehensively surveyed by Mohan~\cite{moh}, the optimization of small moduli sets for balanced hardware implementation remains an underexplored area. Previous investigations by Bajard et al.~\cite{baj_kai_pla,baj_fuk_kiy_pla_sip_sus,baj_mel_pla,baj_fuk_pla} have primarily focused on identifying co-primes within predefined number sets, discovering moduli with limited Hamming weight, and developing sets with specific cryptographic properties. However, the systematic generation of moduli sets that simultaneously maximize dynamic range and minimize bit-width variance has not been adequately addressed.

The alternative paradigm, which forms the foundation of this work work, utilizes a large number of small moduli. This approach offers superior flexibility in system configuration and enables more balanced parallel processing. However, it introduces the challenge of reverse conversion, i.e. transforming residues back to binary representation requires handling intermediate values on the order of $P$.

Recent studies have demonstrated that specialized hardware algorithms enable modular arithmetic implementations on FPGA \cite{gor_sou} and ASIC \cite{gor_vil,gor_vil_2} platforms that achieve up to 30-fold efficiency improvements over conventional EDA-synthesized designs for small-bitwidth moduli (up to 12 bits). This promising result motivates the search for co-prime moduli sets that can provide large dynamic ranges of hundreds or even thousands of bits while utilizing only small-bitwidth moduli. Such an approach would allow RNS to be considered not merely as a computational framework for a limited class of cryptographic applications, but as a viable alternative to conventional positional arithmetic for a broader range of computational tasks.

This work introduces the tool based on a deterministic algorithm for generating optimal co-prime moduli sets within user-specified ranges \cite{coprime_moduli_set}, subject to the following design objectives:
\begin{enumerate}
    \item maximization of the achievable dynamic range $P = \prod_{i=1}^{n} p_i$;
    \item minimization of moduli bit-width variance to facilitate balanced parallel processing and preferential allocation of moduli at the maximum feasible bit-length within the specified range, i.e. the range under consideration must contain as many modules as possible with the highest bit depth from the given range and the greatest number of modules must be of the same bit depth;
    \item mandatory inclusion of a power-of-two modulus to simplify certain RNS operations, i.e. a number of $2^n$ should be within the input range.
\end{enumerate}
While MATLAB's numerical limitations affect dynamic range exceed 1000 bits, the generated moduli sets remain valid and can be verified using alternative tools.

\section{Algorithmic Framework}

Given a range $[X, Y]$ where $X, Y \in \mathbb{N}$ and $X < Y$, the objective is to identify a set $P = \{p_1, p_2, \ldots, p_n\}$ of pairwise co-prime integers satisfying:
\begin{align*}
&\gcd(p_i, p_j) = 1 \quad \forall i \neq j, \\
&X \leq p_i \leq Y \quad \forall i \in \{1, \ldots, n\}, \\
&\exists k : p_k = 2^m \text{ for some } m \in \mathbb{N}, \\
&P = \prod_{i=1}^{n} p_i \text{ is maximized}. 
\end{align*}

The implementation comprises three fundamental modules.

\subsection*{Module 1. Prime Factorization Module (\texttt{multipliers.m})}

This module computes the prime factorization of an input integer $n$, returning the set of distinct prime factors:
\[
Q(n) = \{q_1, q_2, \ldots, q_k\}
\]
where each $q_i$ is a prime number that divides $n$. These factors satisfy the fundamental theorem of arithmetic: there exist unique positive integer exponents $e_1, e_2, \ldots, e_k$ such that
\[
n = \prod_{i=1}^{k} q_i^{e_i}.
\]
Note that the module returns only the set of distinct primes $Q(n)$, not the exponents $e_i$, since for co-primality testing only the presence of common prime factors is relevant.

The algorithm employs trial division with optimization for early termination. For each candidate divisor $d$ from 2 to $\sqrt{n}$, the algorithm tests divisibility and extracts all occurrences of $d$ before proceeding to the next candidate. The computational complexity is $O(\sqrt{n})$ in the worst case, which occurs when $n$ is prime: in this scenario, no divisor is found until the algorithm exhausts all candidates up to $\sqrt{n}$, ultimately returning $Q(n) = \{n\}$. For composite numbers, the early detection of factors and subsequent reduction of $n$ leads to significantly better practical performance.
\\

\textbf{Module 2. Co-primality Testing Module (\texttt{coprimes.m})}

Given two integers $a$ and $b$, this module determines their co-primality by examining the intersection of their prime factor sets. The function returns unity if $\gcd(a,b) = 1$ and zero otherwise. The implementation leverages the factorizations computed by the \texttt{multipliers} module:
\begin{equation*}
\gcd(a,b) = 1 \iff \text{factors}(a) \cap \text{factors}(b) = \{1\}.
\end{equation*}
This approach avoids explicit GCD computation while providing the same correctness guarantee.
\\

\textbf{Module 3. Optimal Set Generation Module (\texttt{copvector.m})}

The core algorithm implements a multi-stage greedy optimization strategy. The procedure can be conceptually decomposed into the following phases:

\textbf{Phase 1: Prime candidate identification.} The algorithm initiates by scanning the range $[X, Y]$ in descending order to identify prime numbers and prime powers. This prioritization ensures that larger primes, which contribute more significantly to the dynamic range, are considered first. Numbers with exactly two factors (unity and themselves) are marked as co-prime candidates.

\textbf{Phase 2: Power-of-two filtering.} If the range contains multiple powers of two ($2^k$ for various $k$), the algorithm selects the largest such power and eliminates all even composite numbers. This strategic choice maximizes both the contribution to the dynamic range and the utility for binary-to-RNS conversion operations.
%

\textbf{Phase 3: Greedy co-prime selection.} The algorithm iteratively constructs the co-prime set by examining candidates in descending order of magnitude. For each candidate, the algorithm verifies co-primality against all previously selected moduli. The selection criterion incorporates a tiebreaking mechanism: when choosing between candidates with identical factor counts, preference is given to the candidate with larger factor product, as this typically indicates a number closer to be a co-prime to the others.

\textbf{Phase 4: Substitution optimization.} After initial selection, the algorithm performs a refinement pass to identify opportunities for improving the dynamic range through substitution. For each candidate $c$ not initially selected, the algorithm computes the product of all selected moduli with which the candidate $c$ shares common factors. If this product is less than the candidate, substituting the candidate for these moduli yields a net gain in the dynamic range:
\begin{equation*}
\text{if } c > \prod_{p_i \in P: \gcd(c,p_i)>1} p_i \text{ then substitute}.
\end{equation*}

\textbf{Phase 5: Dynamic range computation.} The final stage computes the achievable dynamic range. The bit-width required to represent any integer in the range $[0, P - 1]$, where $P = \prod_{i=1}^{n} p_i$, is computed using the fundamental formula:
\begin{equation*}
B_{\text{total}} = \left\lfloor \log_2(P) \right\rfloor + 1 = \left\lfloor \sum_{i=1}^{n} \log_2(p_i) \right\rfloor + 1.
\end{equation*}
For practical implementation, when $P$ can be computed directly (typically when $B_{\text{total}} < 1000$ bits), the algorithm evaluates $P$ and determines its binary width. For extremely large products exceeding MATLAB's numerical precision, the algorithm employs a segmented multiplication approach: the moduli are grouped into segments whose products remain below the 1000-bit threshold, the bit-widths of these segments are accumulated, and a correction term accounting for segment boundaries is applied. This segmentation strategy circumvents MATLAB's data type limitations while maintaining accuracy for dynamic ranges up to approximately $2^{10000}$.

\section{Computational Complexity}

We assume that basic operations (addition, multiplication, comparison) take constant time for numbers up to Y. The implementation consists of three modules.

\textbf{Module 1: multipliers(n).} Computes prime factorization of $n$ using trial division with complexity $O(n)$ per invocation.

\textbf{Module 2: coprimes(a,b).} Determines if $a$ and $b$ are co-prime by comparing their prime factors, requiring two calls to \texttt{multipliers} with total complexity $O(a + b)$ per invocation.

\textbf{Module 3: copvector(X,Y).} Five-phase algorithm:
\begin{enumerate}
    \item Prime identification: $O(N^2 \cdot Y)$;
    \item Even elimination: $O(N^2)$;
    \item Pairwise refinement: $O(k^2 \cdot Y)$;
    \item Composite replacement: $O(k^3 \cdot Y)$ (the most expensive calculation);
    \item Dynamic range computation: $O(k)$,
\end{enumerate}
where $N = Y - X + 1$ is the range size, $Y$ — upper bound, and $k$ is the number of generated co-prime moduli.

Summing all phases and retaining dominant terms:
\begin{align*}
T(X,Y) &= O(N^2 \cdot Y) + O(N^2) + O(k^2 \cdot Y) + O(k^3 \cdot Y) + O(k) =\\
       &O(Y \cdot (N^2 + k^2 + k^3)) + O(N^2 + k).
\end{align*}

Since $k^3 \gg k^2$ for large $k$ and the terms without $Y$ are dominated by those with $Y$ (as $Y \geq 2$), the overall time complexity for arbitrary range $[X, Y]$ is:

\begin{equation*}
T(X, Y) = O\left(Y \cdot (N^2 + k^3)\right).
\end{equation*}

Table~\ref{tab:complexity} presents complexity estimates for various ranges using experimentally generated moduli sets.

\begin{table}[h]
\centering
\caption{Computational complexity for various ranges}
\label{tab:complexity}
\small
\begin{tabular}{l|c|c|c|c}
\hline
Range $[X, Y]$ & $N$ & $Y$ & $k$ & $T(X, Y)$ - number of calculations \\
\hline
\multicolumn{5}{l}{\textit{Wide ranges from 2}} \\
\hline
$[2, 32]$    & 31   & 32   & 11  &  73\ 344 \\
$[2, 64]$    & 63   & 64   & 18  &  627\ 264 \\
$[2, 128]$   & 127  & 128  & 31  &  5\ 877 \ 760 \\
$[2, 256]$   & 255  & 256  & 54  &  56\ 957\ 184 \\
$[2, 512]$   & 511  & 512  & 99  &  630\ 407\ 680 \\
$[2, 1024]$  & 1023 & 1024 & 172 &  6\ 282 \ 265\ 664 \\
$[2, 2048]$  & 2047 & 2048 & 309 &  68\ 988\ 812\ 224 \\
\hline
\multicolumn{5}{l}{\textit{Narrow ranges in upper domain}} \\
\hline
$[17, 31]$   & 15   & 31   & 7   &  17\ 608 \\
$[33, 64]$   & 32   & 64   & 11  &  150\ 720 \\
$[65, 128]$  & 64   & 128  & 17  &  1\ 153\ 152 \\
$[129, 256]$ & 128  & 256  & 27  &  9\ 233\ 152 \\
\hline
\end{tabular}
\end{table}

The algorithm exhibits polynomial complexity with dominant terms from Phase 1 ($N^2 \cdot Y$) and Phase 4 ($k^3 \cdot Y$). Both terms contribute significantly, with their relative importance depending on the specific range characteristics.

\section{Representative Test Cases}

To validate the algorithm and demonstrate its practical applicability, we present several representative co-prime moduli sets generated for different input ranges. 
\subsection*{Example 1: co-primes moduli set from 2 to 32}

Range: from X=2 to Y=32 \\ The number of co-primes in the set is k=11 \\ The dynamic range is 48 bits 
\\ The set of co-primes:
    32    31    29    27    25    23    19    17    13    11
     7
\subsection*{Example 2: co-primes moduli set from 2 to 64}
Range: from X=2 to Y=64 \\
The number of co-primes in the set is k=18\\
The dynamic range is 90 bits \\
The set of co-primes:
    64    61    59    53    49    47    43    41    37    31
    29    27    25    23    19    17    13    11
\subsection*{Example 3: co-primes moduli set from 2 to 128}
Range: from X=2 to Y=128\\
The number of co-primes in the set is k=31\\
The dynamic range is 184 bits \\
The set of co-primes:
   128   127   125   121   113   109   107   103   101    97
    89    83    81    79    73    71    67    61    59    53
    49    47    43    41    37    31    29    23    19    17
    13
\subsection*{Example 4: co-primes moduli set from 33 to 128}
Range: from X=33 to Y=128\\
The number of co-primes in the set is k=25\\
The dynamic range is 157 bits \\
The set of co-primes:
   128   127   125   121   113   109   107   103   101    97
    89    83    81    79    73    71    67    61    59    53
    49    47    43    41    37

\subsection*{Example 5: co-primes moduli set from 65 to 128}
Range: from X=65 to Y=128\\
The number of co-primes in the set is k=17\\
The dynamic range is 113 bits \\
The set of co-primes:
   128   127   125   121   113   109   107   103   101    97
    89    83    81    79    73    71    67

\subsection*{Example 6: co-primes moduli set from 2 to 256}
Range: from X=2 to Y=256\\
The number of co-primes in the set is k=54\\
The dynamic range is 363 bits \\
The set of co-primes:
   256   251   243   241   239   233   229   227   223   211
   199   197   193   191   181   179   173   169   167   163
   157   151   149   139   137   131   127   125   121   113
   109   107   103   101    97    89    83    79    73    71
    67    61    59    53    49    47    43    41    37    31
    29    23    19    17

\subsection*{Example 7: co-primes moduli set from 129 to 256}
Range: from X=129 to Y=256\\
The number of co-primes in the set is k=27\\
The dynamic range is 205 bits \\
The set of co-primes:
   256   255   251   241   239   233   229   227   223   217
   211   199   197   193   191   181   179   173   169   167
   163   157   151   149   139   137   131

\section{Conclusion}

This technical note presented the tool \cite{coprime_moduli_set} for generating optimal co-prime moduli sets  that maximize dynamic range within specified bit-width constraints. The five-phase greedy optimization procedure efficiently identifies sets incorporating both prime and composite numbers while maintaining pairwise co-primality.

A MATLAB-based tool implementing the proposed algorithm has been developed, consisting of the following files:
\begin{itemize}
\item \texttt{copvector.m} --- main algorithm implementation;
\item \texttt{coprimes.m} --- co-primality verification function;
\item \texttt{multipliers.m} --- prime factorization function;
\item \texttt{main.exe} --- standalone executable (requires MATLAB Runtime).
\end{itemize}

The standalone executable \texttt{main.exe} requires MATLAB Runtime (MCR) installation to run on systems without MATLAB. The tool accepts a range $[X, Y]$ as input and outputs the optimal co-prime moduli set with corresponding dynamic range characteristics.

The resulting output file (\textit{coprimes\_result\_X\_Y.txt}) contains the following information: the specified range, the number of co-primes, the dynamic range, and the optimal set of co-primes. The executable file is compatible with Windows.


\end{document}